\documentclass[12pt]{article}
\usepackage{amssymb}

\newcommand{\be}{\begin{equation}}
\newcommand{\ee}{\end{equation}}
\def\bea{\begin{eqnarray}}
\def\eea{\end{eqnarray}}

\newcommand{\bn}{\begin{eqnarray}}
\newcommand{\en}{\end{eqnarray}}

\newcommand{\eps}{\epsilon}

\newcommand{\od}{{\overline{d}}}
\newcommand{\of}{{\overline{f}}}

\newcommand{\p}{\partial}

\newcommand{\nn}{\nonumber}

\newcommand{\no}{\noindent}
\newcommand{\tf}{\tilde{f}}
\newcommand{\td}{\tilde{d}}

\newcommand{\s}{\,\,\,\,}
\def\bea{\begin{eqnarray}}
\def\eea{\end{eqnarray}}

\newcommand{\beq}{\begin{eqnarray}}
\newcommand{\eeq}{\end{eqnarray}}
\begin{document}

\title{\textbf{Embeddings of the ``New Massive Gravity''}}
\author{D. Dalmazi\footnote{dalmazi@feg.unesp.br}, E.L. Mendon\c ca\footnote{eliasleite@feg.unesp.br} \\
\textit{{UNESP - Campus de Guaratinguet\'a - DFQ} }\\
\textit{{Avenida Dr. Ariberto Pereira da Cunha, 333} }\\
\textit{{CEP 12516-410 - Guaratinguet\'a - SP - Brazil.} }\\}
\date{\today}
\maketitle

\begin{abstract}

Using different types of embeddings of equations of motion we investigate the existence of
generalizations of the ``New Massive Gravity'' (NMG) model with the same particle content (massive gravitons).
By using the Weyl symmetry as a guiding principle for the embeddings we show that the Noether gauge embedding approach leads us to a sixth order model in derivatives with either a massive or a massless ghost. If the Weyl symmetry is implemented by means of a Stueckelberg field we obtain a new scalar-tensor model for massive gravitons. It is ghost free and Weyl invariant at linearized level. The model can be nonlinearly completed into a scalar field coupled to the NMG theory. The elimination of the scalar field leads to a nonlocal modification of the NMG.

We also prove to all orders in derivatives  that there is no local, ghost free embedding of the linearized NMG equations of motion around Minkowski space when written in terms of one symmetric tensor.  Regarding that point, NMG differs from the Fierz-Pauli theory, since in later case we can replace the Einstein-Hilbert action by specific $f(R,\Box\, R)$ generalizations and still keep the theory ghost free at linearized level.

\end{abstract}

\newpage

\section{ Introduction}

Massive spin-2 particles can be covariantly described by means of a symmetric rank-2 tensor.  Although this not
the only possible tensor structure, it is very convenient. It is closely connected with a geometrical point of
view (fluctuation about some metric) and it is a minimal description in the sense that we need just one
auxiliary field, i.e., the trace of the tensor which vanishes on shell. If we further require a second-order
theory, in derivatives, we end up with a unique answer: the Fierz-Pauli (FP) theory \cite{fp}. Almost all
developments in massive gravity, from earlier works \cite{vdv,zak,bd,vain} until recent developments, see
\cite{hinter,drham} for review articles, are built up on the top of the FP theory. It is remarkable that absence
of ghosts \cite{bd} and of mass discontinuity \cite{vdv,zak} have been both achieved in recent theories with one
\cite{drg,drgt} and two \cite{hr} dynamic metrics. A good question concerns the uniqueness of those massive
theories, see for instance \cite{hinter2,ky,rmt13,rmt15}. One might speculate on the consequences of abandoning
the FP paradigm and allowing for higher derivative kinetic terms. In fact, the reader can find higher derivative
massive gravities in \cite{no,nos,cds,kno,cs}.

At linearized level, those higher derivative models are specific generalizations of the FP theory which however
share the same spectrum (massive spin-2 particles) without ghosts. This demands of course a more general
investigation of all possible higher derivative generalizations of the FP model.

One way of producing higher derivative models dual to some ``lower derivative'' theory  is by means of the
embedding of its Euler tensor (equations of motion). In particular, this approach can be used to derive the
``New Massive Gravity'' (NMG) of \cite{bht}  from the FP theory as explained in \cite{sd4}. This dualization
procedure can also be used \cite{sd4} to derive the linearized topologically massive gravity (3rd-order in
derivatives) of \cite{djt} from the first-order spin-2 self-dual model of \cite{ak}.

In \cite{euler} we have investigated, to all orders in derivatives, all possible embeddings of the equations of
motion of the FP theory which are ghost free at linearized level. The method consists of adding to the FP theory
quadratic terms in its equations of motion and fix the coefficients of those terms as function of $\Box =
\p_{\mu}\p^{\mu}$ such that at the end we have a dual model with the same particle content of the FP theory. We
have found a system of equations allowing for several solutions for the coefficients. Those theories are in
general of higher order in derivatives but still ghost free at the linearized approximation. They correspond to
modifications of the FP theory in the spin-0 sector of the propagator. They can all be nonlinearly completed
with the help of a fiducial metric. Most of them are $f(R,\Box R)$ modifications of the Einstein-Hilbert kinetic
term plus the FP mass term. Although a complete analysis has not yet been carried out, at least a subset of such
theories can be modified with an appropriate non derivative potential of the type suggested in \cite{drg,drgt}
 and become apparently ghost free beyond the linearized approximation, as shown in \cite{cds}.

On the other hand, to the best we know, there is only one alternative to the FP theory where the spin-2 sector
of the propagator is modified\footnote{In the FP model there is only one massive pole in the spin-2 sector while
in the NMG model there is both a massive and a massless pole, see (\ref{aknmg}), however the massless pole does
not correspond to a propagating particle \cite{oda}.} namely, the $D=3$ NMG theory of \cite{bht}. Here we
investigate the application of embedding procedures on the curvature square NMG theory in order to produce even
higher order models.

The embeddings we have done in \cite{sd4} are all based on requiring some local symmetry, Noether gauge
embedding (NGE), for the dual model while the embedding used in \cite{euler} requires equivalence of the
equations of motion. In section II we first carry out the NGE embedding of the NMG model with respect to the
Weyl symmetry. Next, we introduce a scalar Stueckelberg field and obtain a new scalar-tensor theory for massive
gravitons in $D=2+1$. In section III we look at rather general embeddings where quadratic terms in the NMG
equations of motion are added to the NMG theory with coefficients which are arbitrary functions of $\Box$. We
examine the propagator of the final higher-order model and require equivalence of the particle content, i.e.,
the new dual theory must describe massive spin-2 particles and nothing else. Differently from the FP case we
show here (section III) to all orders in derivatives that there is no local ghost free embedding of the NMG
equations of motion. In particular, the NGE embedding of the Weyl symmetry of section II also leads to a ghost.
The Stueckelberg approach leads to a ghost free model which after elimination of the scalar field becomes
nonlocal. The work in section III is based on the analytic structure of the propagator. In section IV we present
some final comments and our conclusions.

\section{Weyl embedments of the ``New Massive Gravity''}

By a systematic Lagrangian procedure, called Noether gauge embedment (NGE), one can deduce a gauge invariant
massive theory out of a non gauge invariant one \cite{anacleto}. The gauge symmetry of part of the initial
Lagrangian is extended to the whole final theory. However, there is no guarantee that the particle spectrum is
preserved. In \cite{sd4} we have shown that the linearized NMG theory can be obtained via NGE from the usual
Fierz-Pauli theory via embedding of linearized reparametrizations. We have also shown that a linearized higher
derivative topologically massive gravity is obtained from the usual linearized topologically massive gravity of
\cite{djt} via NGE of the Weyl symmetry. In all those cases the particle content is preserved. One could wonder
what would be the gauge invariant action obtained from the linearized New Massive Gravity theory. Since part of
the action of NMG, the curvature square term, is invariant under Weyl transformation
$\delta_{W}h_{\mu\nu}=\phi\eta_{\mu\nu}$, one might try to embed this symmetry into  a new theory. This is what
we next do. The linearized NMG theory can be written, up to an overall constant, as

\bea S_{NMG}&=& 2\, \int d^3x\,\sqrt{-g} \left[- R + \frac{1}{m^2}
\left(R_{\mu\nu}R^{\mu\nu}-\frac{3}{8}R^2\right)\right]_{hh}\nn\\
&=&\int d^3x\,\left[h_{\mu\nu}G^{\mu\nu}(h)+\frac{2}{m^2}G_{\mu\nu}(h)S^{\mu\nu}(h)\right] \label{nmgel}\eea

\no where $G_{\mu\nu}(h)$ is the usual linearized self-adjoint Einstein tensor and $S^{\mu\nu}(h)$ is the
linearized Schouten tensor in $D=3$ defined as $\left\lbrack R_{\mu\nu}(h)-\eta_{\mu\nu}R(h)/4
\right\rbrack_{hh}$ which has the useful property of ``commutativity'' with the Einstein tensor in the sense
that inside integrals $G_{\mu\nu}(h)S^{\mu\nu}(f)=G_{\mu\nu}(f)S^{\mu\nu}(h)$. In the NGE procedure an important
ingredient is the Euler tensor :

\bea K^{\mu\nu} &\equiv & \frac{\delta S_{NMG}}{\delta h_{\mu\nu}} =  2\, G^{\mu\nu}(h)+\frac{4}{m^2}G^{\mu\nu}[S(h)] \nn\\
&=& - \Box \, h^{\mu\nu} - \p^{\mu}\p^{\nu} h + \p^{\mu}\p_{\alpha}h^{\alpha\nu} +
\p^{\nu}\p_{\alpha}h^{\alpha\mu} - \eta^{\mu\nu}[\p_{\alpha}\p_{\beta}h^{\alpha\beta} - \Box \, h ] \nn\\
&-& \frac {\Box}{m^2} \left\lbrack\Box h^{\mu\nu} - \p^{\mu}\p_{\alpha}h^{\alpha\nu} -
\p^{\nu}\p_{\alpha}h^{\alpha\mu} + \frac{\p^{\mu}\p^{\nu} h}2 +
\frac{\eta^{\mu\nu}}2(\p_{\alpha}\p_{\beta}h^{\alpha\beta}-\Box
\, h )\right\rbrack \nn\\
&+& \frac{1}{2\, m^2}\p^{\mu}\p^{\nu}(\p_{\alpha}\p_{\beta}h^{\alpha\beta}) \quad . \label{kg} \eea

\no With the help of an auxiliary field $a_{\mu\nu}$ such that $\delta_{W}a_{\mu\nu}=-\delta_{W}h_{\mu\nu}$ we
implement a first iteration of the form:

\be S^{(1)}=S_{NMG}+\int d^3x\,\,a_{\mu\nu}K^{\mu\nu}.\label{vari}\ee

\no The Weyl variation of (\ref{vari}) can be written as

\be \delta_{W}S^{(1)}=-\int d^3x\,\,\delta_{W}[a_{\mu\nu}G^{\mu\nu}(a)].\ee

\no Therefore we end up with the Weyl invariant theory

\be S_W = S_{NMG} + \int d^3x \left( a_{\mu\nu}K^{\mu\nu} + a_{\mu\nu}G^{\mu\nu}(a) \right) \label{sw1} \ee

\no Noticing that the Euler tensor (\ref{kg}) can be written in terms of the Einstein tensor i.e.: $K^{\mu\nu}=
2 G^{\mu\nu}(H)$ with $H^{\mu\nu} \equiv h^{\mu\nu}+ 2\, S^{\mu\nu}(h)/m^2$  we can rewrite $S_W$ as

\bea S_W &=& S_{NMG} + \int d^3x \left( 2\, a_{\mu\nu}G^{\mu\nu}(H) + a_{\mu\nu}G^{\mu\nu}(a) \right) \nn\\
 &=& S_{NMG} + \int d^3x \left( - H_{\mu\nu}G^{\mu\nu}(H) + (a+H)_{\mu\nu}G^{\mu\nu}(a + H) \right) \quad .
 \label{sw2} \eea

 \no After the shift $a_{\mu\nu} \to \tilde{a}_{\mu\nu} - H_{\mu\nu} $ in (\ref{sw2}), the $\tilde{a}_{\mu\nu}$ auxiliary field
 decouples. We can safely discard the last term $\tilde{a}_{\mu\nu}G^{\mu\nu}(\tilde{a}) $ which is a linearized
 Einstein-Hilbert term without particle content. Thus we have a 6th order Weyl invariant action which turns out
 to have a nonlinear completion,

\bea S_W &=& \int d^3x\,\left\lbrace h_{\mu\nu}G^{\mu\nu}(h)+\frac{2}{m^2}G_{\mu\nu}(h)S^{\mu\nu}(h) -
\left\lbrack h + \frac 2{m^2} S (h)\right\rbrack_{\mu\nu} G^{\mu\nu} \left\lbrack h + \frac 2{m^2} S
(h)\right\rbrack \right\rbrace \nn\\
&=&-\frac{2}{m^2}\int d^{3}x\,\,\left[ G_{\mu\nu}(h)S^{\mu\nu}(h)+\frac{2}{m^2}S_{\mu\nu}(h)G^{\mu\nu}[S(h)]\right]\nn\\
&=& \frac{2}{m^4}\int d^3x \, \left[ \sqrt{-g}
\left(R_{\mu\nu}-\frac{3}{8}g_{\mu\nu}R\right)(\Box-m^2)R^{\mu\nu} \right]_{hh} \quad . \label{sw4} \eea

\no Since the tensor structure of (\ref{sw4}) is the same of the curvature square term of the NMG theory, it is
clear that $S_W$ is invariant under Weyl transformations. The particle content of $S_W$ will be examined in the
next section. The theory $S_W$ contains a ghost.

Another way to embed the Weyl symmetry in the ``New Massive Gravity'' is to introduce a scalar Stueckelberg
field in (\ref{nmgel})  by substituting $h_{\mu\nu} \to h_{\mu\nu} + \eta_{\mu\nu} \, \phi $. Since the fourth
order term of $S_{NMG}$ is Weyl invariant we end up with

 \be S_{\phi}^L[h,\phi] = \int d^3x\,\left[h_{\mu\nu}G^{\mu\nu}(h)+\frac{2}{m^2}G_{\mu\nu}(h)S^{\mu\nu}(h) + \frac 12 \phi \, \Box \, \phi + \frac 12 \phi \left( \Box \, h -
\p^{\mu}\p^{\nu} h_{\mu\nu} \right) \right] \, , \label{nmglphi}\ee

\no By construction, the linear theory $S_{\phi}^L[h,\phi ]$ is invariant under linearized Weyl transformations:
$\delta_W h_{\mu\nu} = \eta_{\mu\nu} \Lambda $ ; $ \delta_W \phi = - \Lambda $. We can easily find a nonlinear
version of $S_{\phi}^L[h,\phi]$, namely,

\be S^{NL}_{\phi}[g_{\mu\nu},\phi]= 2\, \int d^3x\,\sqrt{-g} \left[- R + \frac{1}{m^2}
\left(R_{\mu\nu}R^{\mu\nu}-\frac{3}{8}R^2\right) + \frac 12 \phi \, \Box \, \phi - \frac 12  \phi \, R \right]
\label{nmgenl}\ee

\no It is clear that the equations of motion $\delta \, S^{NL}_{\phi} =0 $ contain the trivial solution
$g_{\mu\nu} = \eta_{\mu\nu}  \, , \, \phi =0 $.  Expanding about such vacuum up until quadratic terms
 in the fluctuations we recover $S^{L}_{\phi}$.
 Therefore, the particle content of $S^{NL}_{\phi}$ consists, at tree level, of one massive spin-2 particle
 just like the NMG of \cite{bht}. However, as in the K-model (massless limit of NMG) studied in detail
 in \cite{deserprl} and \cite{deg}, we might have problems at nonlinear level since the linearized Weyl symmetry
 is probably broken at nonlinear level and consequently the scalar field stops being pure gauge in the full
 model (\ref{nmgenl}). In particular, the phenomenon of bifurcation of constraints found in \cite{deg} might also
 be present here. A detailed study of the constraints structure should be carried out.

\section{Generalized Euler tensor embedment of ``New Massive Gravity''}

Our starting point is the linearized NMG theory with the addition of quadratic terms in its equations of motion:

\be {\cal L}_G [h_{\mu\nu}] = \frac 12 h_{\mu\nu}K^{\mu\nu}+ \frac 12 K_{\mu\nu} \, d(\Box) \, K^{\mu\nu} +
\frac 12 K \, f(\Box) \, K  \label{lg} \ee

\no The first term in (\ref{lg}) is the linearized NMG theory. The NMG Euler tensor $K^{\mu\nu}$ is given in
(\ref{kg}). The coefficients $d(\Box)$ and $f(\Box)$ are so far arbitrary functions of $\Box = \p_{\mu}\p^{\mu}$
such that the Lagrangian ${\cal L}_G $ remains local. Due to the conservation law $\p^{\mu}K_{\mu\nu} = 0$ which
holds identically due to the linearized reparametrization invariance of the NMG ($\delta h_{\mu\nu} =
\p_{\mu}\xi_{\nu} + \p_{\nu}\xi_{\mu} $), other terms which might show up in (\ref{lg}) like
$(\p_{\mu}K^{\mu\nu})^2 $ and $ \p_{\mu}K^{\mu\nu}\p_{\nu}K $ do not contribute\footnote{We have constrained us
here to parity invariant theories, thus avoiding terms like
$\epsilon^{\mu\nu\alpha}K_{\mu\gamma}\p_{\nu}K_{\alpha}^{\,\, \gamma} $.}. In terms of the original field
$h_{\mu\nu}$ we can rewrite ${\cal L}_G$ as

\be {\cal L}_G =  \p^{\mu}h_{\mu\nu} \, c_1(\Box) \, \p_{\alpha}h^{\alpha\nu} + \p^{\mu} h \, c_2(\Box)\,
\p^{\nu}h_{\nu\mu} + h \, c_3(\Box) \, h + h^{\mu\nu} c_4(\Box) \, h_{\mu\nu} + \p^{\mu}\p^{\nu}h_{\mu\nu}
c_5(\Box) \p^{\alpha}\p^{\beta}h_{\alpha\beta} \nn \\ . \label{lp} \ee

\no The coefficients $c_i(\Box)$ are given by\footnote{Henceforth, unless otherwise stated, we replace
$d(\Box),f(\Box))$ by $d,f$ respectively, though they are still arbitrary functions of $\Box$.}

\bea c_1 &=&  \Box \, d -1 + \frac{\Box}{m^2}\left\lbrack 1- 2 \, \Box \, d + \frac{\Box^2}{m^2} \, d
\right\rbrack
\quad , \label{c1} \\
c_2 &=& 1 - \frac{\Box}{2\, m^2} + \Box \, f + \frac{\Box \, d}2 - \frac{\Box \, d}2 \left(\frac{\Box}{m^2} -
1\right)^2  \quad , \label{c2} \\
c_3 &=& \frac{\Box}2 c_2 \quad , \label{c3} \\
c_4 &=& \frac{\Box}2 c_1 \quad , \label{c4} \\
c_5 &=& \frac {f+d}2 + \frac {1}{4\, m^2} + \frac {d\, \Box}{4\, m^2} \left( \frac{\Box}{m^2} - 2 \right) \quad
. \label{c5} \eea

The Lagrangian ${\cal L}_G $ can be further rewritten in terms of a four indices differential operator ${\cal
L}_G  = h^{\mu\nu}G_{\mu\nu\alpha\beta}h^{\alpha\beta} $ whose inverse $G^{-1}$ does not exist due to the
linearized reparametrization symmetry. We choose the de Donder gauge fixing term:

\be {\cal L}_{GF} = \lambda \left( \p^{\mu}h_{\mu\nu} - \p_{\nu}h/2 \right)^2  \quad , \label{gf1} \ee

\no which amounts to shift $ c_1 \to c_1 + \lambda $ , $c_2 \to c_2 - \lambda$ in formulas (\ref{c1}) and
(\ref{c2}) but not in (\ref{c3}) and (\ref{c4}). In (\ref{c3}) we make $c_3 \to c_3 - \lambda \Box/4 $. Before
we display $G^{-1}$ we take a closer look at the coefficients $c_i(\Box )$ in (\ref{c1})-(\ref{c5}). A local
Lagrangian density in terms of $h_{\mu\nu}$  demands

\bea d &=& \frac a{\Box} + \td (\Box) \label{od} \\
f &=& - \frac a{\Box} + \tf (\Box) \label{of} \eea

\no where $\td (\Box)$ and $\tf (\Box)$ are analytic functions of $\Box$ while $a$ is an arbitrary real
constant. In terms of the spin-s projection operators $P_{IJ}^{(s)}$ given in the appendix A and suppressing the
four indices we have

\bea G^{-1} &=& \frac{2\, m^4 \, P_{SS}^{(2)}}{\Box(\Box - m^2)\left\lbrack m^2(1-a) + a\, \Box + \Box (\Box
-m^2)\, \td \right\rbrack } - \frac{2\, P_{SS}^{(1)}}{\lambda \, \Box } + \frac{2 \, P_{SS}^{(0)}}{\Box
\left\lbrack 1 - a +
\Box \, (\td + 2 \, \tf) \right\rbrack } \nn \\
&+& \frac{2 \sqrt{2} \left( P_{SW}^{(0)} + P_{WS}^{(0)} \right)}{\Box \left\lbrack 1 - a + \Box \, (\td + 2 \,
\tf) \right\rbrack } - 4 \frac{\left\lbrack 1 - a + \Box \, (\td + 2 \, \tf) - \lambda \right\rbrack
P_{WW}^{(0)} }{\lambda \, \Box \left\lbrack 1 - a + \Box \, (\td + 2 \, \tf) \right\rbrack } \quad ,
\label{gmenos1} \eea

\no Regarding the local symmetries of ${\cal L}_G$ there is one  special case :

\be a=1 \quad  {\rm and} \quad  \td (\Box) = - 2\, \tf (\Box)  \quad , \label{weylc} \ee

\no since we have a zero in the denominator of the spin-0 sector which indicates a spin-0 symmetry. Indeed,
under a Weyl transformation $\delta_W h_{\mu\nu} = \Lambda \, \eta_{\mu\nu} $ we have from (\ref{lg}) after
integrations by parts

\bea \delta_W {\cal L}_G &=& h ( 2 \, c_4 + 6 \, c_3 - \Box \, c_2 )\Lambda + \p^{\mu}\p^{\nu}h_{\mu\nu} [ 2\,
\Box
\, c_5 - 2\, c_1 - 3 \, c_2 ] \Lambda  \nn\\
&=& (\Box \, h - \p^{\mu}\p^{\nu}h_{\mu\nu} )[1 - a + \Box \, (\td + 2 \, \tf)]\Lambda  \quad . \label{deltawl}
\eea

\no In the special case (\ref{weylc})  we need to add another (Weyl) gauge fixing term, we may choose $ {\cal
L}_{GF}^W = \xi \, h^2 $. This implies the shift $c_3 \to c_3 + \xi$. Consequently, in the Weyl symmetric case
we have

\bea G^{-1}_W &=& \frac{2\, m^4 \, P_{SS}^{(2)}}{\Box^2 ( \Box - m^2 ) [1 - 2\, \of (\Box - m^2)]} - \frac{2\,
P_{SS}^{(1)}}{\lambda \, \Box} + \left( \frac 1{8\xi} - \frac 1{2\, \lambda \, \Box} \right) P_{SS}^{(0)} \nn \\
&+& \left( \frac 1{4\xi} - \frac 1{\lambda \, \Box} \right) P_{WW}^{(0)} + \frac{\sqrt{2}}8 + \left( \frac
1{\xi} + \frac 4{ \lambda \, \Box} \right)\left\lbrack P_{SW}^{(0)} + P_{WS}^{(0)} \right\rbrack \quad .
\label{gmenos1w} \eea

Next we analyze the particle content of ${\cal L}_G$ from the analytic structure of $G^{-1}$ and $G^{-1}_W$. In
momentum space we can calculate the gauge invariant two point amplitude ${\cal A}(k)$ by saturating $G^{-1}$ or
$G^{-1}_W$  with external sources. For instance,

\be {\cal A}(k) = - \frac i2 T_{\mu\nu}^*(k) (G^{-1})^{\mu\nu\alpha\beta}(k) T_{\alpha\beta}(k) \quad .
\label{ak} \ee

\no Where $G^{-1}(k) = G^{-1}(\p_{\mu} \to i \, k_{\mu} )$. Due to the linearized reparametrization symmetry,
the source must be transverse $k^{\mu}T_{\mu\nu} = 0 $, consequently,

\bea {\cal A}(k) &=& i \left\lbrack \frac{S^{(0)}}{k^2 \, [1-a - k^2 ( \od  + 2 \, \of  )]} - \frac{m^4 \,
S^{(2)}}{k^2(k^2 + m^2) [m^2(1-a) - a\, k^2 + k^2 \, \od (k^2 + m^2)]}
\right\rbrack \, , \nn \\
&\equiv & i \left\lbrack \frac{S^{(0)}}{k^2 \, P(k^2)} - \frac{m^4 \, S^{(2)}}{k^2(k^2 + m^2) Q(k^2)}
\right\rbrack \quad , \label{ak2} \eea

\no where $\od = \td (-k^2)$ and $\of = \tf (-k^2)$  are analytic functions of $k^2=k_{\mu}k^{\mu}$ and

\bea
S^{(0)} &=& T_{\mu\nu}^* (P_{SS}^{(0)})^{\mu\nu\alpha\beta} T_{\alpha\beta} = \frac{\vert T \vert^2}2  \quad , \label{s0} \\
S^{(2)} &=& T_{\mu\nu}^* (P_{SS}^{(2)})^{\mu\nu\alpha\beta} T_{\alpha\beta} = T_{\mu\nu}^*T^{\mu\nu} -
\frac{\vert T \vert^2}2 \quad , \label{s2} \eea

\no The quantity $T=\eta_{\mu\nu}T^{\mu\nu} = - T_{00} + T_{ii} $ is the trace of the external source in
momentum space. If the two conditions for Weyl invariance (\ref{weylc}) hold we must have $T=0$.

A key role is played by  the imaginary part of the residue of ${\cal A}(k)$ at each pole. For instance, at  $k^2
= - m^2$ we have:

\be I_m \equiv \Im \, \lim_{k^2 \to - m^2} (k^2 + m^2){\cal A}(k) \quad . \label{im} \ee

\no If and only if $I_m > 0 $ we have a physical particle. If $I_m = 0$ we have a non propagating mode while
$I_m < 0 $ or no definite sign for $I_m$ signalizes the presence of ghost. In order to verify the sign of $I_m $
we fix a convenient coordinate frame splitting the cases of massless and massive poles. In the massless case we
fix a frame such that $k^{\mu} = (k_0,\epsilon,k_0) $, thus, $k^{\mu}k_{\mu} = \epsilon^2$. We will take
$\epsilon \to 0$ at the end. This caution is necessary for the analysis of double poles. From the three
conditions $k^{\mu}T_{\mu\nu} = 0$ we have in this frame

\bea T_{01} &=& - T_{12} - \frac{\epsilon}{k_0} \, T_{11} \quad , \label{t01} \\
T_{02} &=& -T_{22} -  \frac{\epsilon}{k_0} \, T_{12} \quad , \label{t02} \\
T_{00} &=& T_{22}  + 2  \frac{\epsilon}{k_0} \, T_{12} + \frac{\epsilon^2}{k_0^2} \, T_{11} \quad , \label{t00}
\eea

Consequently,

\bea T_{\mu\nu}^* T^{\mu\nu} &=& \vert T_{11}\vert^2 - 2 \frac{\eps}{k_0} (T_{12} T_{11}^* + T_{12}^* T_{11})
\nn\\ &+& \frac{\epsilon^2}{k_0^2} \left\lbrack 2 \vert T_{12} \vert^2 + (T_{11}T_{22}^* + T_{11}^*T_{22}) - 2\,
\vert T_{11}\vert^2 \right\rbrack \, . \label{tmn0} \eea

\be \vert T \vert^2 = \vert T_{11}\vert^2 - 2 \frac{\eps}{k_0} (T_{12} T_{11}^* + T_{12}^* T_{11}) + 2
\frac{\eps^2}{k_0^2} (2\, \vert T_{12}\vert^2 - \vert T_{11} \vert^2 ) \\ \label{t0} \ee

\no In the case of massive poles we choose the frame $k^{\mu}=(m,\eps,0)$ such that $k^2 + m^2 = \eps^2 $. From
$k^{\mu}T_{\mu\nu} = 0 $ we have

\be T_{01} = - \frac{\eps}{m}T_{11} \quad ; \quad T_{02} =  - \frac{\eps}{m}T_{12} \quad ; \quad T_{00} =
\frac{\eps^2}{m^2}T_{11} \quad . \label{t01m} \ee

\no Thus,

\be T_{\mu\nu}^* T^{\mu\nu} = \vert T_{22}\vert^2 + \vert T_{11} \vert^2 \left(1- \frac{\eps^2}{m^2} \right)^2 +
2 \, \vert T_{12} \vert^2 \left(1- \frac{\eps^2}{m^2} \right) \quad , \label{tmnm} \ee

\be  \vert T \vert^2 = \vert T_{22}\vert^2 + \vert T_{11} \vert^2 \left(1- \frac{\eps^2}{m^2} \right)^2 +
\left(1- \frac{\eps^2}{m^2}\right) (T_{11}T_{22}^* + T_{11}^*T_{22}) \quad , \label{tm} \ee

Since $\od (k^2)$ and $\of (k^2)$ are arbitrary analytic functions, there might be double poles in the
denominator of ${\cal A}(k)$. We first examine those poles. From (\ref{s0}),(\ref{s2}),(\ref{tmnm}) and
(\ref{tm}) we see that is impossible to take linear combinations of $S^{(0)}$ and $S^{(2)}$ in order to end up
only with terms of order $\eps^2$. Therefore a massive double pole $1/(k^2 + m^2)^{2} = 1/\eps^{4}$ can not be
reduced to a simple pole by any fine tuning of the functions $\od $ and $\of$. So henceforth we assume that all
massive poles must be simple poles. The conclusion remains the same for the Weyl symmetric case. In the later
case $T_{00} = T_{11} + T_{22}$ and (\ref{t01m}) imply $T_{22} = - T_{11} + (\eps^2/m^2)T_{11} $ which does not
help canceling the term  $\vert T_{12} \vert^2 $ in (\ref{tmnm}).

The massless case is a bit different. From (\ref{s0}),(\ref{s2}),(\ref{tmn0}) and (\ref{t0}) we see that we do
have one special combination of order $\eps^2$ which may turn double poles into simple ones, namely,

\be S^{(2)} -  S^{(0)} = T_{\mu\nu}^* T^{\mu\nu} - \vert T \vert^2 = \frac{\eps^2}{k_0^2}\left\lbrack
(T_{11}T_{22}^* + T_{11}^*T_{22}) - 2 \, \vert T_{12} \vert^2 \right\rbrack\quad . \label{zero} \ee

\no However, the term $(T_{11}T_{22}^* + T_{11}^*T_{22})$ has no definite sign. So we end up with a ghost. In
the special case of Weyl symmetry, using (\ref{t00}) in  $T_{00}=T_{11} + T_{22}$ we have  $T_{11} =
2(\eps/k_0)T_{12} + {\cal O}(\eps^2) $. Thus, the dangerous term of (\ref{zero}) becomes of order $\eps^3$ and
will not contribute to the residue. So we may hope to turn a double massless pole $1/k^4$ into a a physical pole
only in the Weyl invariant case after a specific fine tuning of $\od$ and $\of$. In particular, this is the
mechanism behind the fourth order K-term which describes a physical massless particle as explained in
\cite{deserprl} via decomposition of $h_{\mu\nu}$ in orthogonal modes and in \cite{unitary} via analytic
structure of the propagator.

In summary, multiple poles lead us to ghosts in general except for the  double massless pole in the Weyl
symmetric case which will be examined later on.

Henceforth we split our analysis in four cases:

\vskip .7 cm

\no \leftline{$a\ne 1$ ({\it Case} I)} \leftline{$a=1$ and $\od + 2\, \of \ne 0 $ ({\it Case} II)}
\leftline{$a=1$ and $\od + 2\, \of = 0 $ ({\it Case} III )} \leftline{$a=0=\od $ and $\of = 1/(2\, k^2) $ ({\it
Case} IV)}

\no In the cases III and IV we have Weyl symmetry. The case IV corresponds to the $S_{\phi}$ model of
(\ref{nmglphi}) after elimination of $\phi $.

\subsection{{\it Case} I : $a\ne 1$}

As a warm up  we start reproducing the results of \cite{oda} for the NMG theory. We take $a=0=\od=\of$. We have
one massless and one massive pole,

\be {\cal A} (k) = i \left\lbrack \frac{S^{(0)}}{k^2} - \frac{m^2 \, S^{(2)}}{k^2(k^2+m^2)}\right\rbrack \quad .
\label{aknmg} \ee

\no Taking $\eps \to 0$ in (\ref{tmn0}) and (\ref{t0}) we have a vanishing residue at the massless pole and
consequently a non propagating mode:

\be I_0 = \Im \lim_{k^2 \to 0} k^2 \, {\cal A}(k) =  S^{(0)} - S^{(2)} = \frac{\vert T \vert^2}2 -
\left(T_{\mu\nu}^*T^{\mu\nu} - \frac{\vert T \vert^2}2\right) = 0 \, , \label{i0nmg} \ee

\no The residue at the massless pole vanishes for the very same reason as it does in the Maxwell-Chern-Simons
theory of \cite{djt}, namely, the lowest order term (in derivatives) of the theory (linearized Einstein-Hilbert)
has no particle content. Taking $\eps \to 0$ in (\ref{tmnm}) and (\ref{tm}) we have a positive residue at the
massive pole, a physical massive spin-2 particle,

\be I_m  = \Im \lim_{k^2 \to -m^2} (k^2 + m^2) \, {\cal A}(k) = S^{(2)} = 2\, \vert T_{12} \vert^2 + \frac{\vert
T_{11}- T_{22} \vert^2}2 > 0 \quad . \label{imnmg} \ee

Now we go back to the general case $a \ne 1$. Except for the NMG case $a=0=\od=\of$ which will not be treated
here anymore, we have in general extra massive poles stemming from the polynomial $Q(k^2)$. Requiring that no
tachyons show up we can write

\be {\cal A} (k) = \frac{-i\, m^2 \, S_A (k^2) \,  \prod_{i=1}^{N_Q} m_i^2}{(a-1)k^2(k^2+m^2)(k^2 + m_1^2)\cdots
(k^2+m_{N_Q}^2)} \quad , \label{sa} \ee

\no where $N_Q \ge 1$ is the number of extra massive poles coming from $Q(k^2)$. Since we are specially
interested in the massive poles, we can write from (\ref{s0}),(\ref{s2}) and  (\ref{tmnm}), (\ref{tm}) at $\eps
\to 0$,

\be S_A (k^2) \equiv S^{(2)} + A(k^2) \, S^{(0)} = 2 \vert T_{12} \vert^2 + \vert T_{11} \vert^2 + \vert T_{22}
\vert^2 + \frac{(A-1)}2 \vert T_{11} + T_{22} \vert^2 \quad . \label{Ak} \ee

\no The quantity $A(k^2)$ is an analytic real function of $k^2$ whose specific form is not important, it is
defined by comparing (\ref{sa}) with (\ref{ak2}). Defining the polynomial of degree $N_Q + 1$:

\be {\cal P}(k^2) = (k^2+m_0^2)(k^2 + m_1^2)\cdots (k^2+m_{N_Q}^2) \quad , \label{px} \ee

\no where $m_0^2 \equiv m^2$ is the mass squared already present in the NMG theory, it is clear that the sign of
the residue $I_{m_j}$ at some pole $k^2 = - m_j^2$ depends essentially upon the sign of the ratio $S_A/{\cal
P}^{\prime}$ calculated at $k^2 = -m_j^2$, where ${\cal P}^{\prime}= d{\cal P}/dk^2$. Since the derivative of a
polynomial has alternating signs at its consecutive simple zeros, the only hope of having positive residues at
the different massive poles is to require that $S_A$ also has alternating signs at such points. However, it is
easy to prove that $S_A(-m_j^2)$ either has no definite sign or is definite positive. The point is that, if
$A(-m_j^2) \ge 0$ we can guarantee that the last three terms of (\ref{Ak}) add up to a non negative number, so
in those cases $S_A(-m_j^2) \ge 0 $. On the other hand, since the three complex numbers $T_{12},T_{11},T_{22}$
are totally unconstrained, even if we take $A(-m_j^2) < 0 $, depending on the relative strength of those three
complex numbers, the sign of (\ref{Ak}) may change. Thus, we can not guarantee that $S_A(-m_j^2) < 0 $. In
conclusion, whenever we have more than one massive pole we have ghosts and only the NMG case is safe at $a \ne
1$.

\subsection{{\it Case} II : $a = 1$ and $\od + 2\, \of \ne 0 $}

In this case we have in principle a double massless pole and massive poles:

\be {\cal A} (k) = -i \left\lbrace \frac{m^4 S^{(2)}}{k^4(k^2+m^2)[(k^2+m^2)\od - 1]} + \frac{S^{(0)}}{k^4(\od +
2\, \of )} \right\rbrace \quad . \label{ak3} \ee

\no  We can choose $\od (0) + \of (0) = \frac 1{2\, m^2}$ and  turn the double massless pole into a simple one.
However, since we have no Weyl symmetry, as explained in the paragraph of formula (\ref{zero}), we are doomed to
have a massless ghost. Therefore we go to the next case.

\subsection{{\it Case} III : $a = 1$ and $\od + 2\, \of = 0 $  }

If we set $a=1$ and $\td = -2\, \tf $, or $\od = -2\, \of $, in formulas (\ref{c1})-(\ref{c5}) and plug those
results in (\ref{lp}) we have

\bea {\cal L}_W &=& \frac 1{m^4}\p_{\mu}h^{\mu\nu} \Box \, H(\Box) \p^{\alpha}h_{\alpha\nu} - \frac 1{2\, m^4}
\p^{\mu}h \Box \, H(\Box)
\p^{\alpha}h_{\alpha\mu} - \frac 1{4\, m^4} h\, \Box^2 \, H(\Box) h \nn \\
&+& \frac 1{2\, m^4} h_{\mu\nu}\Box^2 \, H(\Box) h^{\mu\nu} + \frac 1{4\, m^4} \p^{\mu}\p^{\nu}h_{\mu\nu} \,
H(\Box) \p^{\alpha}\p^{\beta}h_{\alpha\beta} \quad , \label{lw} \eea

\no where $H(\Box) = (\Box - m^2)[ 1- 2\, (\Box - m^2)\tf (\Box) ] $. The above Lagrangian can be nonlinearly
completed in terms of square of curvatures in the form of a K-term of the NMG theory, i.e.,

\be {\cal L}_W^{NL} = \frac{2\sqrt{-g}}{m^4}\left( R^{\mu\nu}- \frac 38 g^{\mu\nu}\, R \right)(\Box - m^2)[ 1-
2\, (\Box - m^2)\tf (\Box) ] R_{\mu\nu} \quad , \label{lwnl} \ee

\no In the case $\tf =0$ we recover the Weyl embedding of the last section, see (\ref{sw4}).

Back to the two-point amplitude  ${\cal A}(k)$, we have again a double massless pole and massive poles in
general. Assuming that the analytic function $\of (k^2) = \tf (\Box \to - k^2 )$ is such that we have no
tachyons, we can write

\be {\cal A}(k) = i \frac{m^4 S^{(2)}}{k^4(k^2+m^2)[1 + 2\, \of (k^2 + m^2)]} = i \frac{m^2
S^{(2)}\prod_{i=0}^{N_Q}m_i^2}{k^4 {\cal P}(k^2)} \quad . \label{ak4} \ee

\no where ${\cal P}(k^2)$ is defined in (\ref{px}).  From (\ref{s2}), (\ref{tmnm}) and the fact that $T=0$ due
to the Weyl symmetry, we have at each massive pole :

\be I_{m_j} = \Im \, \lim_{k^2 \to - m_j^2} (k^2 + m_j^2){\cal A}(k) = \frac{2(\vert T_{12} \vert^2 + \vert
T_{11} \vert^2 )}{{\cal P}^{\prime}(-m_j^2)} \quad . \label{imj2} \ee

\no Due to the alternating signs of ${\cal P}^{\prime}$ at its consecutive single zeros, it is impossible to
have $I_{m_j} > 0 $ for all $j=0,\cdots, N_Q$. We are forced to assume $N_Q=0$, i.e., $\of = 0$. In this subcase
${\cal P}(k^2)=k^2 + m^2 $, so ${\cal P}^{\prime} = 1$ and the massive pole is a physical one $I_{m}=I_{m_0}
=2(\vert T_{12} \vert^2 + \vert T_{11} \vert^2 )>0 $. Regarding the massless double pole, we have already seen
that due to the Weyl symmetry we have $T_{11} = 2(\eps/k_0)T_{12} + {\cal O}(\eps^2) $, substituting back in
(\ref{tmn0}) we obtain \cite{unitary}

\be T^{\mu\nu}T^*_{\mu\nu} = - 2 \frac{\epsilon^2}{k_0^2}\vert T_{12} \vert^2 + {\cal O}(\eps^3) \quad .
\label{tmn2} \ee

\no Consequently, although the apparent double pole has become a simple pole we still have a ghost due to the
negative sign of the residue:

\be I_{0} = \Im \, \lim_{k^2 \to 0} k^2 {\cal A}(k) = \lim_{\eps \to 0} \frac{
m^2}{\eps^2}T^{\mu\nu}T^*_{\mu\nu} = - \frac{2\, m^2}{k_0^2} \vert T_{12} \vert^2 < 0 \quad , \label{i02} \ee

\no Therefore, the Weyl invariant theory (\ref{sw4})  of last section will unavoidably contain a ghost.

\subsection{{\it Case} IV : $a = 0=\od $ and $ \of = 1/(2 \, k^2) $  }

In this last case we have a nonlocal theory corresponding to the Weyl invariant action $S_{\phi}$ given in
(\ref{nmglphi}) after the elimination of $\phi$:

\be S^{L}_{\phi}[g_{\mu\nu},\phi]= 2\, \int d^3x\,\sqrt{-g} \left[- R + \frac{1}{m^2}
\left(R_{\mu\nu}R^{\mu\nu}-\frac{3}{8}R^2\right) - \frac 18 R\, \frac 1{\Box} \, R \right]_{hh}
\label{nmgenl}\ee

\no After adding a gauge fixing term like (\ref{gf1}) plus another one for the Weyl symmetry ${\cal L}_{GF}^W =
\zeta \, h^2 $ the action acquires the form
 (\ref{lp}) with the coefficients:

 \be c_1 = \frac{\Box}{m^2} - 1 + \lambda \quad ; \quad c_2 = \frac 12 - \frac{\Box}{2m^2} - \lambda \quad ; \quad c_3 = - \frac{\Box^2}{4\, m^2} + \frac{\Box^2}{4}(1-c) \quad , \nn \ee
 \be c_4 = \frac{\Box}{2 m^2}(\Box - m^2) \quad ; \quad  c_5 = \frac 1{4\, m^2\Box}(\Box - m^2) \quad . \label{coef} \ee

\no The propagator, suppressing indices,  is given by

\be G^{-1} =  \frac{2\, m^2 \, P_{SS}^{(2)}}{\Box(\Box - m^2)} - \frac{2\, P_{SS}^{(1)}}{\lambda \, \Box } +
\frac{ ( P_{WW}^{(0)} +  P_{SS}^{(0)})}{8}\left(\frac 1\zeta - \frac 4{\lambda\, \Box}\right) + \frac{(4 \zeta +
\lambda\, \Box ) }{4 \zeta\lambda \Box}  \left( P_{SW}^{(0)} + P_{WS}^{(0)} \right)  \ee

\no After saturating the propagator with transverse and traceless sources as in (\ref{ak}) we are left with two
simple poles, one massive and one massless which come both from the pure spin-2 sector:

\be {\cal A}(k) = - i \frac{m^2 \, S^{(2)} }{k^2 (k^2 + m^2)} \quad . \label{akst} \ee

\no As expected, the dependance on the gauge parameters $\lambda$ and $ \zeta$ disappear which guarantees gauge
invariance of the two-point amplitude. When we look closer at the massless pole using $k_{\mu} =
(k_0,\epsilon,k_0)$,  it is easy to see that its residue vanishes with power $\epsilon^2$. From (\ref{tmn0}),
(\ref{t0}) and
 $T=\eta_{\mu\nu}T^{\mu\nu} =0 $ we have

\be I_0 = \Im \lim_{k^2 \to 0} k^2 \, {\cal A}(k)  = -  \lim_{\epsilon \to 0} T_{\mu\nu}^*T^{\mu\nu} =
-\lim_{\epsilon \to 0} \frac{\epsilon^2}{k_0^2} \left\lbrack -2 \vert T_{12} \vert^2 + (T_{11}T_{22}^* +
T_{11}^*T_{22})\right\rbrack = 0 \label{i0} \ee

\no The massive pole is a physical one (positive residue). From (\ref{tmnm}) we have

\be I_m  = \Im \lim_{k^2 \to -m^2} (k^2 + m^2) \, {\cal A}(k) = S^{(2)} = \vert T_{22} \vert^2 + 2 \vert T_{12}
\vert^2  > 0 \label{im} \ee

\no Therefore, the linearized version of the  model (\ref{nmgenl}) is unitary and contains only massive
gravitons in the spectrum.

\section{Conclusion}

There are several higher-order, in derivatives, modifications of the Fierz-Pauli theory which describe massive
spin-2 particles and are still ghost free at linear level. At nonlinear level they can be identified with
$f(R,\Box \, R)$ modifications of the Einstein-Hilbert theory plus the Fierz-Pauli (FP) mass term, see
\cite{euler}. All such modifications of the FP model occur in the spin-0 sector of the theory. Some of  those
models have been further changed, see e.g. \cite{cs}, by the addition of a convenient non derivative nonlinear
potential of the type found in \cite{drg,drgt} in order to account for the absence of ghosts at nonlinear level.
They define massive gravity theories with interesting cosmological properties.

Another alternative massive spin-2 theory is the so called ``New Massive Gravity'' (NMG) \cite{bht} which exists
only in $D=3$. Its propagator differs from the FP case also in the spin-2 sector. In section III we have
generalized the embedding procedure used in \cite{euler} in order to search for arbitrary higher-order
modifications of the NMG model. Contrary to the FP case, we conclude that there is no local ghost-free
modification of the NMG theory. We have carried out a thorough calculation of the residues at all possible
massive and massless poles in the propagator. There is always a pole with negative residue (ghost).

In section II we have used the Weyl symmetry as a guiding principle for the embedding of the NMG. First we have
looked at the Noether gauge embedment of the Weyl symmetry, which leads to (\ref{sw4}). In this case we only
have one massive and one massless pole. Unfortunately, their residues have opposite signs. If we reverse the
overall sign of (\ref{sw4}), we have a physical massless graviton and a massive ghost. So we might hopefully use
the Weyl invariant model (\ref{sw4}) with reversed sign as a phenomenological toy model along the lines of
\cite{ss}. Namely, we can have a consistent unitary theory if the ghost mass stays above the energy cut-off of
the theory. Although the model (\ref{sw4}) is of sixth-order in derivatives, the double massless pole is reduced
to a simple pole and the analytic structure of the propagator is similar to some curvature square modifications
of general relativity in $D=4$.

Still in section II we have obtained a promising candidate for a consistent massive gravity different from the
NMG model. From the introduction of a scalar Stueckelberg field in the linearized version of the NMG theory we
have derived the linearized Weyl invariant model given in (\ref{nmglphi}). As in the case of (\ref{sw4}) we have
one massless and one massive pole but differently from (\ref{sw4}) both poles are simple poles. The Weyl
symmetry now kills the residue at the massless pole such that the ``would be'' massless ghost does not propagate
at all as shown at the end of section III (Case IV). The residue at the massive pole is positive and we are left
with physical massive gravitons, at least in the linearized approximation. The model can be nonlinearly
completed leading to the scalar-tensor theory (\ref{nmgenl}) which might be an alternative to the usual NMG
model. However, as in the case of the pure K-term (massless limit of NMG) analyzed in \cite{deg}, the Weyl
symmetry is probably broken beyond the linearized level which might lead to a ghost in the full theory. This is
probably true also for the higher derivative topologically massive gravity of \cite{sd4,andringa}. Those
examples demand detailed investigations of the constraints structure which are beyond the scope of the present
work.

%


\section{Appendix}

Here we display the operators $P_{IJ}^{(s)}$, the coefficients $A_{ij}(\Box)$. They make use, as building
blocks, of the spin-0 and spin-1 projection operators acting on vector fields, respectively,

\be  \omega_{\mu\nu} = \frac{\p_{\mu}\p_{\nu}}{\Box} \quad , \quad \theta_{\mu\nu} = \eta_{\mu\nu} -
\frac{\p_{\mu}\p_{\nu}}{\Box}\quad , \label{pvectors} \ee

\no  we define the spin-s operators $P_{IJ}^{(s)}$ acting on symmetric rank-2 tensors in $D$ dimensions:

\be \left( P_{SS}^{(2)} \right)^{\lambda\mu}_{\s\s\alpha\beta} = \frac 12 \left(
\theta_{\s\alpha}^{\lambda}\theta^{\mu}_{\s\beta} + \theta_{\s\alpha}^{\mu}\theta^{\lambda}_{\s\beta} \right) -
\frac{\theta^{\lambda\mu} \theta_{\alpha\beta}}{D-1} \quad , \label{ps2} \ee

\be \left( P_{SS}^{(1)} \right)^{\lambda\mu}_{\s\s\alpha\beta} = \frac 12 \left(
\theta_{\s\alpha}^{\lambda}\,\omega^{\mu}_{\s\beta} + \theta_{\s\alpha}^{\mu}\,\omega^{\lambda}_{\s\beta} +
\theta_{\s\beta}^{\lambda}\,\omega^{\mu}_{\s\alpha} + \theta_{\s\beta}^{\mu}\,\omega^{\lambda}_{\s\alpha}
\right) \quad , \label{ps1} \ee

\be \left( P_{SS}^{(0)} \right)^{\lambda\mu}_{\s\s\alpha\beta} = \frac 1{D-1} \,
\theta^{\lambda\mu}\theta_{\alpha\beta} \quad , \quad \left( P_{WW}^{(0)} \right)^{\lambda\mu}_{\s\s\alpha\beta}
= \omega^{\lambda\mu}\omega_{\alpha\beta} \quad , \label{psspww} \ee

\be \left( P_{SW}^{(0)} \right)^{\lambda\mu}_{\s\s\alpha\beta} = \frac 1{\sqrt{D-1}}\,
\theta^{\lambda\mu}\omega_{\alpha\beta} \quad , \quad  \left( P_{WS}^{(0)}
\right)^{\lambda\mu}_{\s\s\alpha\beta} = \frac 1{\sqrt{D-1}}\, \omega^{\lambda\mu}\theta_{\alpha\beta} \quad ,
\label{pswpws} \ee

\no They satisfy the symmetric closure relation

\be \left\lbrack P_{SS}^{(2)} + P_{SS}^{(1)} +  P_{SS}^{(0)} + P_{WW}^{(0)} \right\rbrack_{\mu\nu\alpha\beta} =
\frac{\eta_{\mu\alpha}\eta_{\nu\beta} + \eta_{\mu\beta}\eta_{\nu\alpha}}2 \quad . \label{sym} \ee

\no and the algebra

\be \left(P^{(s)}\right)_{IJ} \left(P^{(r)}\right)^{JK} = \delta^{rs} \left(P^{(s)}\right)_I^{\, \, K} \quad .
\label{algebra} \ee

\section{Acknowledgements}

The works of D.D. and E.L.M. are partially supported by CNPq under grants (307278/2013-1) and (449806/2014-6)
respectively.


\begin{thebibliography}{99}

\bibitem{fp} M. Fierz and W. Pauli, Proc. Roy. Soc. Lond. A {\bf 173} (1939) 211.

\bibitem{vdv}H. van Dam, M.J.G. Veltman,  Nucl.Phys. B22 (1970)
397-411.

\bibitem{zak}V.I. Zakharov, JETP Lett. 12 (1970) 312.

\bibitem{bd}D.G. Boulware and S. Deser,  Phys.Rev. D6 (1972) 3368-3382.

\bibitem{vain}A. I. Vainshtein, Phys. Lett. B 39, 393 (1972).


\bibitem{hinter} K. Hinterbichler,   Rev.Mod.Phys. 84 (2012) 671-710,
see also arXiv:1105.3735.

\bibitem{drham} C. de Rham, ``Massive Gravity'',  arXiv:1401.4173 (2014).


\bibitem{drg}C.de Rham, G. Gabadadze, Phys.Rev. D82 (2010) 044020

\bibitem{drgt}C.de Rham, G. Gabadadze and A. J. Tolley, Phys.Rev.Lett. 106 (2011) 231101.

\bibitem{hr}S.F. Hassan and R. A. Rosen, JHEP 1202 (2012) 126.

\bibitem{hinter2}K. Hinterbichler, JHEP 1310 (2013) 102.

\bibitem{ky}R. Kimura, D. Yamauchi Phys.Rev. D88 (2013) 084025.

\bibitem{rmt13} C. de Rham, A. Matas, A. J. Tolley, Class.Quant.Grav. 31 (2014) 165004.

\bibitem{rmt15} C. de Rham, A. Matas, A. J. Tolley, Class.Quant.Grav. 32 (2015) 21, 215027

\bibitem{no} S. Nojiri, S.D. Odintsov, Phys. Lett. B 716 (2012)
377.

\bibitem{nos} S. Nojiri, S.D. Odintsov, N. Shirai, J. Cosmol. Astropart.
Phys. 1305 (2013) 020.

\bibitem{cds} Yi-Fu Cai, F. Duplessis, E. N. Saridakis
arXiv:1307.7150.

\bibitem{kno} J. Kluson, S. Nojiri and S.D. Odintsov, Phys. Lett. B 726
(2013) 918 [arXiv:1309.2185].

\bibitem{cs} Yi-Fu Cai, E. N. Saridakis,
arXiv:1401.4418.

\bibitem{bht} E. Bergshoeff, O. Hohm and P.K. Townsend, Phys.Rev.Lett. 102:201301,2009.

\bibitem{sd4}D. Dalmazi and E.L. Mendon\c ca, JHEP 0909 (2009)011.

\bibitem{djt} S. Deser, R. Jackiw and S. Templeton, Ann. of Phys. {\bf 140}(1982)
372.

\bibitem{ak} C. Aragone and A. Khoudeir, Phys. Lett.
B{\bf173} 141 (1986).

\bibitem{euler}D. Dalmazi, A.L.R. dos Santos, E.L. Mendon\c ca, Class.Quant.Grav. 32 (2015) 1, 015022.

\bibitem{oda}M. Nakasone, I. Oda, Prog.Theor.Phys. 121 (2009) 1389-1397.

\bibitem{anacleto} M.A. Anacleto, A. Ilha , J.R.S. Nascimento, R.F. Ribeiro, C. Wotzasek,
Phys.Lett.B504:268-274,2001.

\bibitem{deserprl} S. Deser, Phys.Rev.Lett. 103 (2009) 101302.

\bibitem{deg} S. Deser, S. Ertl, D. Grumiller , J.Phys. A46 (2013) 214018.

\bibitem{unitary} D. Dalmazi, Phys.Rev. D80 (2009) 085008.

\bibitem{ss} F. O. Salles and Ilya L. Shapiro,  Phys.Rev. D89 (2014) 8, 084054, see also Phys.Rev. D90 (2014) 12,
129903.

\bibitem{andringa}  R. Andringa, Eric A. Bergshoeff, M. de Roo, O. Hohm, E. Sezgin and P. K. Townsend,
Class.Quant.Grav. {\bf 27} (2010) 025010 , arXiv:0907.4658.




\end{thebibliography}
\end{document}